# How War Distorts International Trade: Gravity-Model Evidence from Europe after the Russia-Ukraine Conflict

**Luigi Capoani, Margarita Shnaider, Piergiorgio Martini***

## Abstract

This paper investigates how geopolitical conflict reshapes trade patterns, focusing on the economic consequences of the Russo-Ukrainian war on European and global trade flows. War is conceptualized as a shock that increases bilateral trade costs within a structural gravity model, rather than as a force acting against trade flows, amplifying frictions in territories closer to the epicenter and reducing the economic attractiveness of major trade routes. The empirical analysis combines an Extended Gravity Model based on bilateral trade data from 2019 and 2023 with geographic, institutional, and political factors, including sanctions regimes and energy specialization. The findings show that war not only reduces trade volumes but also operates multiplicatively on trade frictions, influencing both the intensity and direction of trade disruptions, with more pronounced effects in the central corridors of the European market. As a result, some trade relationships collapse while others are redirected towards less exposed regions; furthermore, policy choices are decisive in shaping trade flows and contribute to isolating the Russian economy by creating a policy-induced trade void around the target country, while mechanisms such as the EU Single Market facilitate the internal reallocation of trade flows, preserving economic cohesion.

*Corresponding Author at: Department of Economics, Ca' Foscari University of Venice, Cannaregio, 873, Venezia, 30100, VE,Italy. E-mail address: luigi.capoani@unive.it; and Department of Physics and Astronomy, University of Padua, Padua, Italy. E-mail address: rita.shnayder@gmail.com and; Department of Business and Management, Luiss Guido Carli University, Rome, Italy and at European Youth Think Tank, 1 Place des Orphelins, Strasbourg, France. E-mail address: piergio.martini@gmail.com

# 1. Introduction

The relationship between war, geopolitical instability, and international trade has become a crucial topic of contemporary economic debate. Conflicts not only affect the macroeconomic fundamentals of the countries involved but also produce disturbances that propagate through the economic space, modifying the architecture of trade flows and the ability of markets to maintain integration and stability. The Russo-Ukrainian war has made this vulnerability particularly evident in the European context, characterized by high geographical interdependence, concentrated production specialization, and strong energy dependence. The impact of the conflict was manifested through increases in energy costs, historically high inflationary pressures, declines in investment and a general deterioration in economic expectations, as shown by recent reports of the European Commission.

The combination of these phenomena has led to the recognition of the need for analytical tools capable of simultaneously capturing the spatial dimension of markets, the institutional structure of economic integration, and the distorting effects resulting from geopolitical shocks. Walter Isard's contribution fits into this context, and his work has provided the basis for a certain reading of the international economy, in which economic interactions take the form of attractive and repulsive forces subject to constraints of distance, economic mass, and political conditions. His extension of the law of gravity to the economic sphere (Isard 1954), as well as the subsequent development of regional science, Peace Science and Peace Economics, forms the theoretical foundation for interpreting how localized shocks—such as wars, political frictions, or diplomatic tensions—alter the potential for trade attraction between countries and redraw trade balances. Isard's gravitational approach allows us to consistently represent how a conflict acts as an anti-gravitational force: it increases effective economic distance, weakens trade ties in areas near the epicenter, and pushes flows toward more distant or less exposed regions. Recent studies confirm this insight. Glick & Taylor (2010), for example, show how wars generate persistent collapses in bilateral trade, while Taralashvili (2024) highlights that even "soft" conflicts—such as diplomatic flare-ups, boycotts and civil protests—exert long-term effects, which typically fade away within five years, reducing trade volumes by approximately 8.6% and weakening regional cohesion. Expanding upon this, Capoani et al. (2025a; 2025b) interpreted Brexit and geopolitical instabilities as perturbations of the Single Market gravitational field, demonstrating that for the same shock, the effects are more intense in regions characterized by greater economic concentration, such as the Blue Banana corridor.

In light of these developments, the Russo-Ukrainian war can be read as a paradigmatic case of anti-gravitational shock: it markedly alters the structure of trade costs, reduces the ability of central markets (such as Germany, France, and the Netherlands) to maintain trade attraction, and introduces persistent frictions that amplify the fragility of a European economy already strained by the pandemic. The data reported by the European Commission for the period 2022-2024 show how rising energy and raw material prices have fueled an inflationary spiral, weakened investment and production, and increased precautionary household savings—all elements that fit coherently into the logic of the field distortions generated by the conflict.

In order to analyse these shocks more effectively, this article will be divided into two complementary sections. The first part provides the theoretical foundation by building on the framework of Walter

Isard's studies, a gravitational simulation model prior to the Russia Ukraine war. The second part adopts an empirical perspective, applying the theoretical framework to assess the economic effects of the armed conflict. It conducts simulations of conflict scenarios using data from European Union databases and introduces an innovative methodology based on the gravitational model of trade.

In addition, other variables will be analyzed alongside gravity to determine which factor best explains and most strongly influences trade.

Finally, it will be seen that certain countries, for instance, the United States and Canada, gain considerably more in trade from the conflict, while others, such as Germany, the Netherlands and Great Britain, experience significant trade losses as a result of the Russia-Ukraine war.

## 2. Literature Review

Isard's intellectual legacy covers three complementary areas. The first is the econophysical gravitational model, widely recognized as a cornerstone of international economics (Deardorff, 1998; Isard, 1954). Building on this foundation, Isard developed the field of regional science and later inspired Peace Science, a research area focused on the causes and resolution of conflicts. Its economic counterpart, Peace Economics, was described by Anderton (2017) as the economic analysis of conflict. Following Isard's vision, it explores how economic instruments can help prevent, mitigate, and measure the effects of conflicts on societies, governments, and markets. From a normative standpoint, it also calls for a reallocation of military spending toward socially productive activities (Isard, 1994). Despite his success in linking trade theory and regional science, however, Isard never fully unified these disciplines with Peace Economics into a single theoretical framework.

In pursuit of this integration, Isard (1988) introduced mathematical and interdisciplinary methods to study the origins and evolution of conflicts. Building on Richardson's arms-race theory, which associates perceived threats with higher levels of armament, he demonstrated that excessive military expenditures reduce national resources and constrain further investment (Richardson, 1960).

Recent evidence from the Russia-Ukraine war confirms these dynamics on a broader scale. The conflict has generated profound economic instability across the European Union, as surging energy and commodity prices have fueled inflation and undermined productive investment. In 2022, inflation reached historic levels—averaging over 9% across the EU—largely due to the war's disruptive effects on energy markets and supply chains (European Commission, 2023). According to Lampa and Garbellini (2022), highly energy-dependent economies such as Germany, Italy, and several Eastern European member states faced potential GDP contractions of up to 4%, illustrating how armed conflicts can destabilize economic growth and jeopardize post-pandemic recovery.
Following the initial shock, a gradual phase of disinflation has taken hold, supported by tighter monetary policies, higher interest rates, and targeted fiscal measures (European Commission, Directorate-General for Economic and Financial Affairs, 2024).
Despite these stabilizing efforts, the macroeconomic landscape across the European Union remains fragile and structurally weakened. This fragility largely stems from heightened uncertainty among consumers and firms, triggered by the inflationary surge in essential goods and energy following the Russia-Ukraine war. Households, confronted with persistent price pressures and elevated borrowing costs, have restrained consumption and increased precautionary savings, while firms have postponed or reduced investment amid tighter financial conditions (Verwey et al., 2024). Consequently, economic growth and productive activity have slowed markedly, reflecting a pervasive climate of stagnation and weak confidence (Mbah & Wasum, 2022).

The Russian-Ukrainian war has not only disrupted short-term performance but has also produced enduring structural distortions with long-lasting implications, amplifying the vulnerabilities of the European economy. The European Central Bank has itself acknowledged that lowering interest rates alone cannot swiftly revive growth or resolve these structural weaknesses (Reuters, 2024), since such transformations require time, sustained investment, and significant financial resources to restore competitiveness and stability (Verwey et al., 2024).

To better contextualize this structural fragility, it is useful to recall the broader post-pandemic trajectory. After the sharp contraction of 2020 and the vigorous rebound of 2021, growth in 2022 was moderate and uneven, constrained by soaring inflation, disruptions in global value chains, and persistent energy insecurity (European Commission, 2023). In the subsequent years, 2023 and 2024, recovery has remained cautious and fragmented, as high input costs, declining real purchasing power, and tight financial conditions have continued to weigh on investment and access to credit (European Commission, Directorate-General for Economic and Financial Affairs, 2024). These adverse dynamics have particularly affected the most vulnerable groups—low- and middle-income households and energy-intensive industries—resulting in a tangible loss of competitiveness and social welfare (European Commission, 2023). Although recent data from the European Commission, Directorate-General for Economic and Financial Affairs (2024), indicate that inflation is gradually converging toward more sustainable levels (around 2–3% within the EU), the economic repercussions of the war remain deeply embedded in the continent's growth trajectory, leaving its economies exposed to persistent fragility, structural imbalances, and renewed geopolitical and energy shocks.

This framework emphasizes the need for a better comprehension of the relationship between trade and conflicts. From here, Isard's studies carry a great explanatory power that can be used for this case. Recent research has further developed Isard's vision by seeking to unify his three main domains: international trade, regional science, and peace economics.

Specifically, Capoani et al. (2025a) applied a gravitational field model to the European Single Market, interpreting Brexit as a perturbation phenomenon within the European Union's commercial sphere. Their findings show that the most destabilizing shocks occur near regions of high economic concentration, such as the European Blue Banana, where intense activity amplifies the effects of conflict and instability. This confirms the gravitational field model as a powerful tool for analyzing the spatial diffusion of trade and geopolitical shocks, where the magnitude of disruption depends on both proximity and intensity. Expanding on this approach, Capoani et al. (2025b) reinterpreted conflicts and sociopolitical instability as negative masses, or anti-gravitational forces that weaken trade attraction and economic cohesion. Their results reveal a strong relationship between a conflict's proximity to the economic core and its disruptive potential, thereby linking Isard's Peace Economics with contemporary trade theory and contributing to a unified framework for assessing the economic consequences of geopolitical instability. To further operationalize this integrated vision, Capoani and Martini (2025c, 2026) empirically tested how geopolitical shocks propagate across economic space. Their study extends the gravitational framework developed in prior works to include real-world conflict dynamics, focusing on the European Union during the 2018–2022 period. By introducing the Russian–Ukrainian war as a natural experiment, they evaluate how spatial proximity to the conflict alters macroeconomic outcomes such as GDP growth and inflation. The analysis distinguishes between pre-war, pandemic, and wartime conditions, allowing the authors to isolate the specific economic disruption generated by the invasion.

Several empirical contributions beyond the authors' previous works further validate the application of gravity-based frameworks to the study of conflicts and trade disruption. Glick and Taylor (2010) developed one of the earliest large-scale analyses of war-related trade shocks using a gravity model applied to 20th-century data. Their research estimated both the immediate and long-term effects of major conflicts on bilateral trade flows, showing that wars produce large and persistent declines in international commerce, even after hostilities formally end. This finding underscores the systemic fragility of global markets in the aftermath of conflict and supports the idea that war-related shocks propagate through the economic system with gravitational asymmetries. Such evidence reinforces the rationale for adopting dynamic gravity-based methodologies in contemporary analyses of the economic consequences of armed conflict.

A more recent study by Taralashvili (2024) expanded this approach by examining how *soft conflicts*– defined as diplomatic flare-ups, disputes and boycotts –affect bilateral trade flows. Using an econometric gravity model, the author demonstrated that even in the absence of direct military confrontation and the imposition of sanctions, such tensions significantly reduce trade volumes between countries (averaging an 8.6% decrease), with effects that persist over time but commonly tail off within five years. The analysis revealed that non-violent geopolitical frictions act as "negative gravitational forces" in the international trade system, weakening regional integration and undermining market cohesion. These results broaden the conceptual foundation of the present study, supporting its aim to capture both hard and soft forms of instability within a unified gravitational framework for analysing economic conflict.

### 2.1 Presentation of the Theoretical Framework

Walter Isard (1954) was the pioneer in adapting the gravitational law from physics to the study of international trade, laying the groundwork for what he termed the theory of income potential. Decades later, Fujita and Krugman characterized their contributions to the New Economic Geography as a continuation of Isard's original vision to unify trade theory with spatial economics. They described their work as "a continuation, perhaps even a validation, of Isard's dream of bringing space back to the heart of economics" (Fujita et al., 2004, p.153), thereby recognizing the profound influence of Isard's seminal texts, Location and Space Economy (1956) and its subsequent elaboration in General Theory (Isard et al., 1969).

Following *Capoani (2023),* the proposed model can be expressed as follows:

$$i1V = \sum_{j=1}^{n} i1V_j = \sum_{j=1}^{n} k \frac{Y_j}{d_{ij}^{a}} \qquad (1)$$

Here, $i1$V represents the potential income that each state generates in relation to state $i$, while $i1V_j$ denotes the potential income that country $j$ produces relative to country $i$. $Y_j$ stands for the national or regional income of country $j$, and $d_{ij}$ indicates the average effective distance between countries $i$ and $j$ (a metric adjusted to reflect transportation costs). The constant $a$ is an exponent applied to $d_{ij}$ and $k$ functions as a constant similar to the one used in the physical law of gravity.

Isard also develops the notion of economic distance to capture how goods move between countries through different transport channels. His model is built on the assumption that nations that are geographically proximate and economically alike tend to engage more intensively in trade, as they encounter lower logistical and communication costs and often share similar institutional and cultural frameworks. This interconnectedness becomes evident when considering the U.S.-led global trading network. For decades, the American presence has sustained a system of open exchange marked by low trade barriers and efficient flows of goods, services, and capital among economically comparable and culturally aligned countries (Mansoor, 2024).

According to Isard, any analytical model that treats distance as a factor contributing to higher social or economic costs must inherently exhibit a gravitational structure. Similar to the physical principle, distance within social or economic systems exerts a comparable influence: the closer two states are, the greater their potential for economic exchange. Hence, a state located next to the country *i* will have the greatest potential income in relation to *i*.

Nevertheless, the concept of distance extends beyond mere geography. It also captures the notion of resistance, a factor that may increase not only with spatial separation but also as a result of socio-political dynamics such as cooperation, rivalry, or conflict among nations. Even when countries are geographically close, political tensions or policy disagreements can function as barriers that widen their effective economic distance. For instance, rising political frictions, such as those following the Brexit referendum, led to heightened policy uncertainty, which reduced the productivity of UK firms and slowed global trade growth (Aiyar et al., 2023). This demonstrates how political conflict can weaken the gravitational pull between neighbouring economies, diminishing their potential for economic interaction despite geographic proximity.

This dynamic is clearly illustrated by the analysis of the U.S.-led global trading system. A political or strategic withdrawal of the United States from international markets would not only reduce its own trade and productivity but also trigger a chain reaction affecting all the economies closely linked to it. Because these nations are economically and institutionally intertwined, the retreat of one major actor disrupts previously established trade flows, weakens mutual interdependence, and widens the effective economic distance between countries that were once strongly connected (Mansoor, 2024).

## 2.2 The Gravitational Field of Trade and the Disruptive Effects of Conflict

The gravitational field model of trade, first introduced by Isard (1954), marks a significant step beyond the traditional bilateral gravity models by conceptualizing international economics as a complex network of interrelated forces rather than as a set of isolated country pairs. While Tinbergen’s early framework (1962), inspired by Newtonian mechanics, viewed trade as a phenomenon occurring between two independent entities, Isard’s approach describes the world economy as a dynamic system of interacting “masses,” where each national economy generates a gravitational field that extends outward and interacts with others. The intensity of this field depends on both the economic size of a country—typically represented by its GDP (Y)—and its spatial distance ($D_{ij}$) from other economies.

This field can be imagined as a set of radiating lines spreading outward, like concentric ripples created when a stone falls into water. For perfectly regular regions, these lines are evenly distributed, while for more irregular or asymmetric territories, they curve and distort. Smaller or less diverse areas tend to create almost circular fields of attraction, whereas larger and economically heterogeneous regions generate irregular, overlapping zones of influence. Just as in physics, the commercial gravitational field theoretically extends to infinity, but its effective pull weakens as distance increases: proximity strengthens attraction, while remoteness diminishes it.

When multiple economies interact, their gravitational fields overlap and combine through a process akin to the principle of scalar and vector addition. This overlap forms a composite gravitational field that represents the overall configuration of global trade flows. Within this composite field, certain regions emerge as centres of gravity (areas of maximum attraction where trade and investment are most concentrated). These centres correspond to zones of strong economic magnetism, functioning as hubs that draw firms, capital, and infrastructure, thereby reinforcing processes of cumulative growth and agglomeration.

Building on this theoretical foundation, Isard's framework can be extended beyond trade analysis to illuminate the dynamics of conflict within the global economy. In this reinterpretation, conflict functions as a *repelling force* that distorts and weakens the gravitational equilibrium of trade fields. During times of peace, border regions often thrive thanks to their accessibility and the ease of cross-border exchanges. However, when conflict arises, these same borders transform into barriers or frontlines, exposing local economies to instability and military disruption. The long-term fate of such regions depends on the conflict's duration, intensity, and resolution: while some may recover and even attract postwar investment through reconstruction, others face prolonged decline and uncertainty. When hostility increases, it acts as an effective expansion of distance, weakening trade ties and reducing the gravitational pull between economies. Conflict can therefore be introduced into the model as a corrective term, which alters the stability of economic fields.

Ultimately, the complete equation in this model encompasses its primary term from the Theory of Commercial Gravitational Fields (Capoani, 2023) –built upon the gravity model applied to international trade introduced by Tinbergen (1962) –combined with the formalization of geopolitical disruptive effects as presented by Capoani and Martini (2026). Therefore, the subsequent equation captures the total economic and geopolitical potential between countries and is expressed as follows:

$$G_{tot} = \left(\sum_{j=1}^{n} K \frac{Y_j}{D_{ij}}\right) - \left(\sum_{h=1}^{m} \frac{conflict\ size\ h}{D_{from\ the\ conflict\ h}}\right) \qquad (2)$$

The first term represents the cumulative attractive potential generated by the economic masses of interacting states, while the second term quantifies the disruptive effect of conflict. The nearer and more intense a conflict is relative to a major economic centre, the stronger its negative impact on the surrounding trade field. This mirrors the physical principle whereby an opposing force introduced near the centre of gravity produces greater destabilization in comparison to one located at the periphery.

From this relationship emerges the concept of *conflict potential*, a spatially defined negative field centred around the origin of hostilities. The closer one moves toward this epicentre, the more trade volumes and economic exchanges decline, ultimately approaching zero at the point of maximum confrontation. The extent of the conflict's influence can be represented as a circular field whose radius corresponds to the degree of disruption.

An illustrative example of this mechanism can be found in the "Blue Banana," Europe's principal economic axis stretching from northern Italy through Germany to southern England. A severe conflict within this core—such as between France and Germany—would substantially distort the gravitational field of the entire European Single Market, generating a cascade of negative effects across the continent. Conversely, a dispute involving more peripheral countries, such as Ireland and the United Kingdom, would exert a weaker influence on the overall system, given their greater geographic and economic distance from the continent's gravitational core.

In conclusion, the gravitational field model of trade not only explains the spatial logic underlying global economic interactions but also offers a powerful framework for understanding how conflict disrupts them. Peace strengthens the gravitational cohesion of markets, fostering convergence and growth, whereas war and instability introduce distortions that repel economic activity, fragment networks, and weaken the structural integrity of the global trade system.

# 3. Data and methodology

Building upon the previously outlined gravitational model of trade, the purpose of this section is to employ both the gravity model and the concept of gravitational fields to show that a conflict occurring near the center of a market (such as the European one) has far greater destabilizing effects than an otherwise identical conflict of equal intensity located in a peripheral zone. To illustrate this dynamic, consider the well-known Blue Banana corridor: a major dispute arising at its core, as a conflict between France and Germany, could severely destabilize the entire European Single Market. Conversely, a confrontation between the United Kingdom and Ireland would likely have a far more limited impact, since these countries occupy a peripheral position relative to Europe’s commercial and geographic core, thereby limiting the systemic repercussions of their dispute.

## 3.1 Data

For our empirical analysis, we rely on the Dynamic Gravity dataset compiled by Gurevich and Herman (2018) and provided by the United States International Trade Commission (USITC).
This dataset is broadly comparable to those produced by Centre d’Études Prospectives et d’Informations Internationales (CEPII) (Conte et al., 2022) or to those constructed by Andrew Rose *(Rose, 2000, 2004),* both of which yield results consistent with standard international trade theory. Nonetheless, while the qualitative trends remain similar across these datasets, the quantitative outcomes may occasionally diverge.

In this study, only the distance variable from the USITC dataset was employed, and its estimates were not statistically different from those derived using CEPII data.
This choice is further supported by the USITC’s documentation, which notes certain limitations in the CEPII dataset. Firstly, its list of included countries is static, failing to account for the emergence or disappearance of states over time, which may introduce bias in evaluating the impact of various gravity-related factors on trade flows. Secondly, several CEPII variables exhibit little variation over time or in proportion to the magnitude of specific conflicts, meaning that the dataset may not fully capture the real intensity of such events. To address these shortcomings, the USITC dataset incorporates the key variables from earlier models but enriches them with additional dimensions, specifically, temporal and scale variations in conflicts, to enhance the precision and transparency of the resulting estimates. It includes a wide range of bilateral and unilateral indicators for 285 countries and territories spanning the period from 1948 to 2016. These include measures of institutional stability, macroeconomic performance, and geographical attributes, selected in line with established principles of international trade theory and conventional gravity model techniques.

The distance variable used in the USITC dataset represents the population-weighted distance between the origin country ($country_o$) and the destination country ($country_d$), calculated using the distances between each country’s principal cities weighted by their respective populations. Distances between a country and itself (for example, Italy-to-Italy) are deliberately excluded from the computation of bilateral trade flows, as the paper focuses on international trade flows between countries, rather than on the measurement of internal trade attraction.

The study centers on the European market, covering 47 countries in Europe, including both EU member states and non-EU economies such as the United Kingdom, Norway, Switzerland, and Turkey. For each country, annual GDP values (in U.S. dollars) for the year 2019 were taken from the World Bank Database. The choice of 2019 as the reference year serves several purposes: it avoids distortions arising from the global disruptions caused by the SARS-CoV-2 pandemic, mitigates the effects of geopolitical instability linked to the war in Ukraine, and addresses the limited availability of reliable post-2019 data.

## 3.2 Methodology

This study employs an extended gravitational model, inspired by Walter Isard's framework, to analyze the economic impact of the Russia-Ukraine conflict on global trade flows. The model conceptualizes the conflict as an anti-gravitational force that repels trade near Ukraine while potentially redirecting it to distant regions.

The analysis was conducted in two phases. During the gravitational simulation model, we compute baseline gravitational forces, apply a conflict-induced shock, or so-called anti-gravitational forces, and visualize the results through a grid-based heatmap. This approach prioritizes geographical distance to Ukraine as the primary explanatory variable to isolate proximity effects. In this phase we use bilateral trade data from the USITC database (2015-2019). In the extended gravitational model we expand our study by building our own gravity dataset starting from 2022 up to 2024 for the verification of the previously built model during the gravitational simulation model.

The following section outlines the data, model formulation, shock mechanism, visualization, and robustness considerations.

### 3.2.1 Gravitational Simulation Model (Pre-Conflict)

The analysis utilizes the USITC dataset, which includes bilateral trade flows, GDP (constant USD millions from Penn World Tables), population-weighted great-circle distances, and country coordinates (latitude/longitude). The dataset is a version of the CEPII Gravity Dataset, which uses population-weighted great-circle distance (in km). It is calculated using population data from major cities to determine the “economic center of gravity" for each country. To ensure data integrity, the dataset is preprocessed to remove rows with missing critical variables, such as distance and GDPs for both the origin and destination countries. Due to the missing values for 2019, data from 2017 were used. Therefore, it should be kept in mind that the so-called economic centers (countries with the biggest GDP), as will be shown later, are defined on the basis of their 2017 economic conditions. However, as the model is primarily theoretical, the analysis focuses on relative changes in the simulations rather than on absolute numerical values.

The baseline gravitational force $F_{norm}$ between origin country (o) and destination country (d) is calculated using Isard's gravity model:

$$F_{norm} = \frac{GDP_o \cdot GDP_d}{dist^2}$$

This formula assumes that trade attraction is proportional to the product of GDPs and inversely proportional to squared distance, mirroring Newton's gravitational law.

To simulate the conflict's disruptive effect, a shocking factor is introduced as an anti-gravitational force that decays with distance from the conflict epicenter, designated as Ukraine in this study. The distance to the focal point is then computed using the Haversine formula for great-circle distances (in kilometers), with coordinates sourced from the dataset as the reference point.

The shocking factor is defined as

$$s = \frac{R^2}{dist^2_{Ukraine} + \varepsilon} \quad (4)$$

where R is the shock radius, and ε is the value that prevents division by zero (1e-10), which produces strong repulsion near Ukraine (e.g., at 100 km, the shocking factor is ~ 30.25) and weaker effects at greater distances. For the shocking radius we have chosen the starting point to be approximately 5 degrees or ~550 km.
The shocked force is computed as

$$F_{shock} = F_{norm} \cdot \left(1 - s_p \cdot s\right) \quad (5)$$

with $s_p$ being a constant parameter shock, which quantifies the conflict intensity ($s_p = 3$), amplifying repulsion (e.g., at 100 km, reduction by ~99%) and boosting distant flows (e.g., ~300% increase). For flows involving Ukraine (where the origin or destination country if the trade is 'Ukraine'), we apply an additional 90% reduction to localize the anti-gravity effect. This creates a balanced simulation: trade "repels" from the conflict zone but "attracts" elsewhere, preventing an unrealistically global reduction in flows.

The analysis of the mean, standard deviation and maximum value of the resulting $F_{shock}$ indicates that the flows near Ukraine are getting reduced, while the force increases in distant flows.

To capture the net effect, the difference force was calculated using the following formula:

$$F_{diff} = F_{norm} - F_{shock} \quad (6)$$

To facilitate visualization, forces are normalized using a log-scaled approach to handle skewness:

$$\underline{F_{norm}} = \frac{loglog\,(1 + F_{norm})}{loglog\,(F_{norm} + \varepsilon)} \quad (7)$$

$$\underline{F_{shock}} = \frac{loglog\,(1+F_{shock})}{loglog\,(\,F_{shock}+\varepsilon)} \qquad (8)$$

This transformation compresses extreme values and rescales them to [0,1], enhancing interpretability despite large magnitude variations.

Countries are filtered to focus on significant trade contributors, defined as those whose $F_{shock}$ share is bigger than 0.00005 relative to total flow. Ukraine is explicitly included using index union, resulting in a dataset of 2,859 bilateral flows across 31 countries, including the United Arab Emirates, Belgium, China, and the United States.

### 3.2.2 Anti-Gravity Effect ($F_{tg}$)

The global anti-gravity effect is quantified as

$$F_{tg} = \sum \langle F_{shock} \rangle \cdot F_{eff} \qquad (9)$$

$$\text{where } \langle F_{shock} \rangle = \frac{F_{shock}}{\sum F_{shock}} \qquad (10)$$

$F_{tg}$. weights flows by their shocked contribution, and $F_{eff}$. is zero except for conflict mask flows, where it equals $F_{diff}$. The local value of the conflict anti-gravity force is 8.58e^05, while the global value F_target = -0.9663 indicates a negative effect driven by Ukraine's trade reductions, though its magnitude reflects the country's relatively small trade share.

### 3.2.3 Results

This analysis examines the impact of the Russia-Ukraine n War on principal European Union trade routes, defined as flows where at least one trading partner is an EU member. The force of these trade connections was calculated using a gravitational Isard model introduced earlier. The standard gravitational model, referred to  in our study as a *normal* state—as shown in (3.5)—was then compared against a *shocked* scenario—as specified in (3.6)—where the conflict introduces a significant new source of friction to the system. In the analysis we assume that the conflict epicenter is Ukraine; therefore, we have calculated the distance from each origin country (e.g., 'Germany-Netherlands route', where the origin country is Germany) to the focal point. After that, we arrange the trade routes from the closest to more distant ones. For representativeness, only routes within a distance of 6000 km were included. In fact, the shocking factor reduces drastically with the increase of the distance and considering the choice of the shocking radius to be around 550 km, the order of magnitude becomes less than $10^{-2}$ . Thus, it would have a negligible impact.

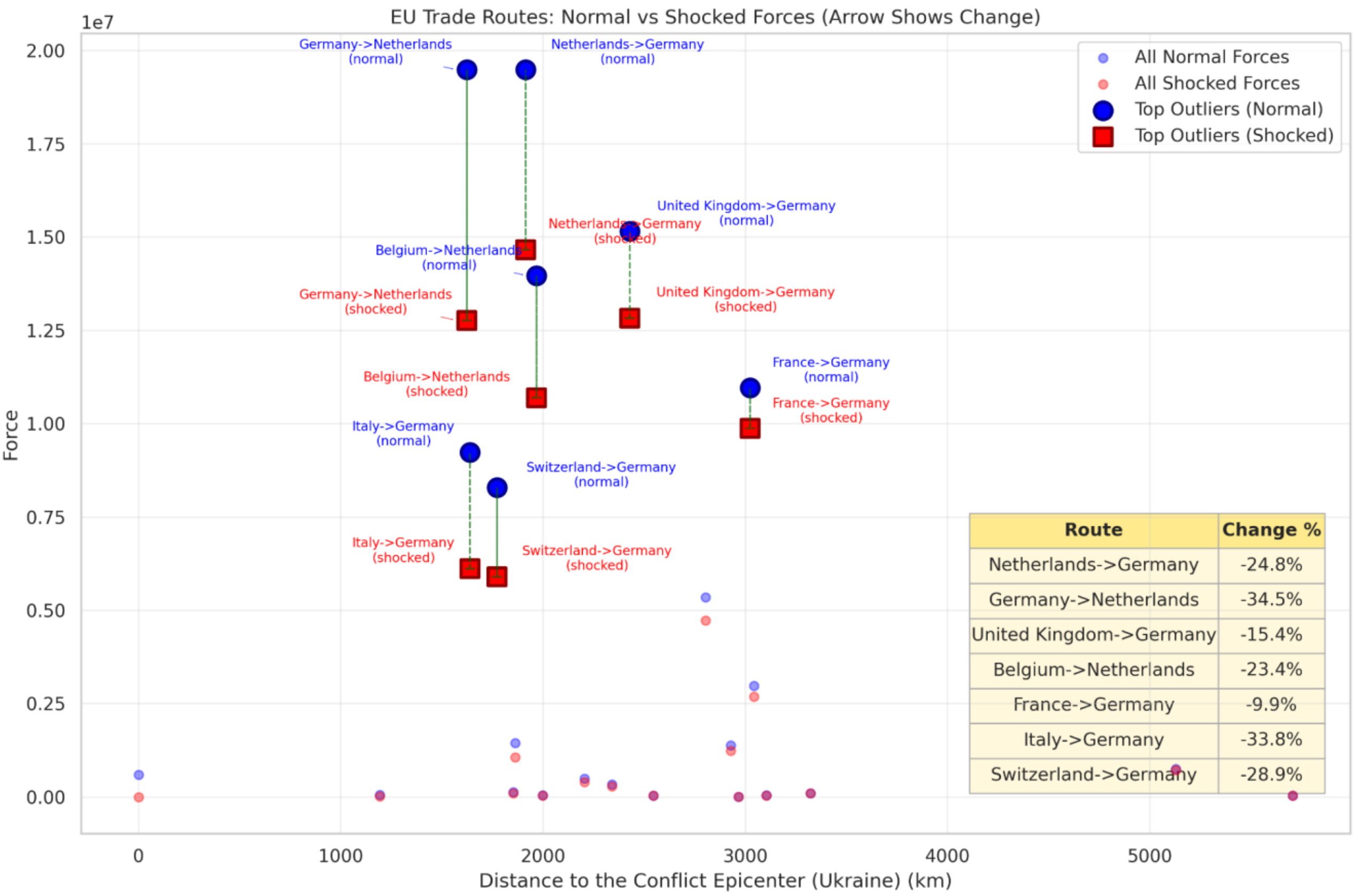

| Route | Change % |
|---|---|
| Netherlands->Germany | -24.8% |
| Germany->Netherlands | -34.5% |
| United Kingdom->Germany | -15.4% |
| Belgium->Netherlands | -23.4% |
| France->Germany | -9.9% |
| Italy->Germany | -33.8% |
| Switzerland->Germany | -28.9% |

*Figure 1: Simulations of trade routes of* the *Top Outliers* with the biggest gravitational force. *Personal work.*

For the first simulation, the routes selected for this analysis were the *Top Outliers* with the biggest gravitational force, representing the most intense and economically critical trade corridors. In fact, Figure 1. illustrates that the primary finding is that the conflict has induced a universally negative shock on the trade forces of Europe's largest economies, yet the severity is highly conditional. As illustrated in the chart, all *Top Outlier* routes (predominantly featuring Germany, the Netherlands, and France) experienced a substantial decrease from their Normal (blue) to Shocked (red) state. This is quantified by reductions ranging from -9.9% with regard to the flows from France to Germany to a severe -34.5% for flows from Germany to the Netherlands. These fluctuations are due to the large baseline “Force”. The initial impact of the Netherlands, thanks to the Rotterdam Port being Germany’s gateway, makes it one of the most important trade routes in Europe. Therefore, even if the geographical proximity to the conflict center would bring the decrease in the trade force up to 100% and more, on the global scale or on the one within Europe, it would not be as prominent.

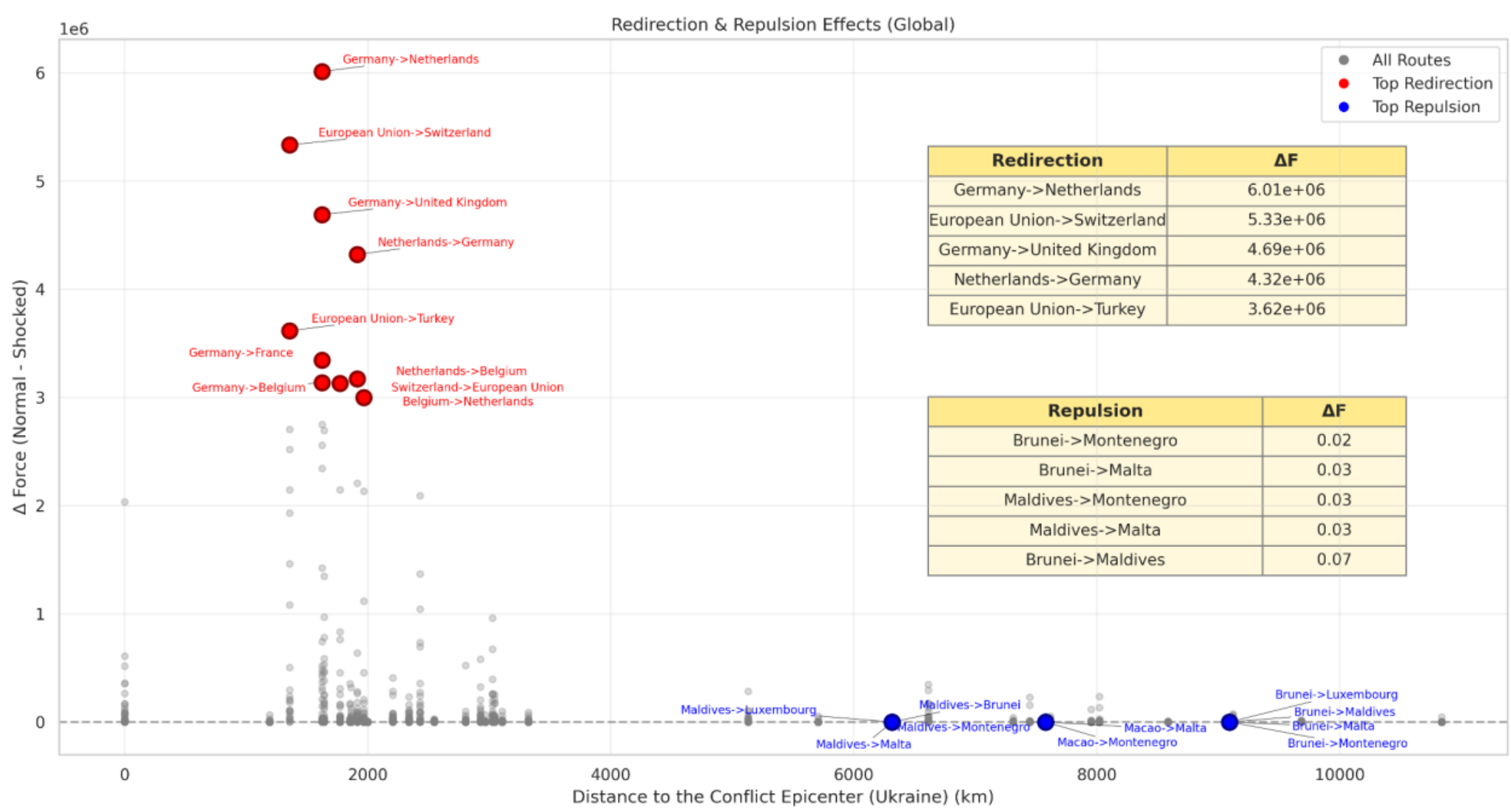

| Redirection | ΔF |
|---|---|
| Germany->Netherlands | 6.01e+06 |
| European Union->Switzerland | 5.33e+06 |
| Germany->United Kingdom | 4.69e+06 |
| Netherlands->Germany | 4.32e+06 |
| European Union->Turkey | 3.62e+06 |

| Repulsion | ΔF |
|---|---|
| Brunei->Montenegro | 0.02 |
| Brunei->Malta | 0.03 |
| Maldives->Montenegro | 0.03 |
| Maldives->Malta | 0.03 |
| Brunei->Maldives | 0.07 |

*Figure 2: Simulations of trade routes in which either the origin or destination country is in the EU. Personal work*

On the other hand, Figure 2 simulates the routes in which either the origin or destination country is in the EU to isolate them from the routes out of interest. This focused analysis reveals that the most significant disruptions (redirection) are concentrated entirely among core, high-intensity intra-European trade corridors. Flows from Germany to the Netherlands and from Germany to the United Kingdom show the largest absolute loss of force, confirming that they are highly impacted due to their combination of high economic importance and relative proximity to the conflict. Conversely, the routes demonstrating "repulsion" (the least effect) are minor routes connecting distant, small economies (e.g., Brunei, Maldives) with smaller EU states like Malta and Luxembourg. Their negligible ΔF is a clear result of their great distance combined with an already low initial trade force, rendering them effectively immune to the shock.

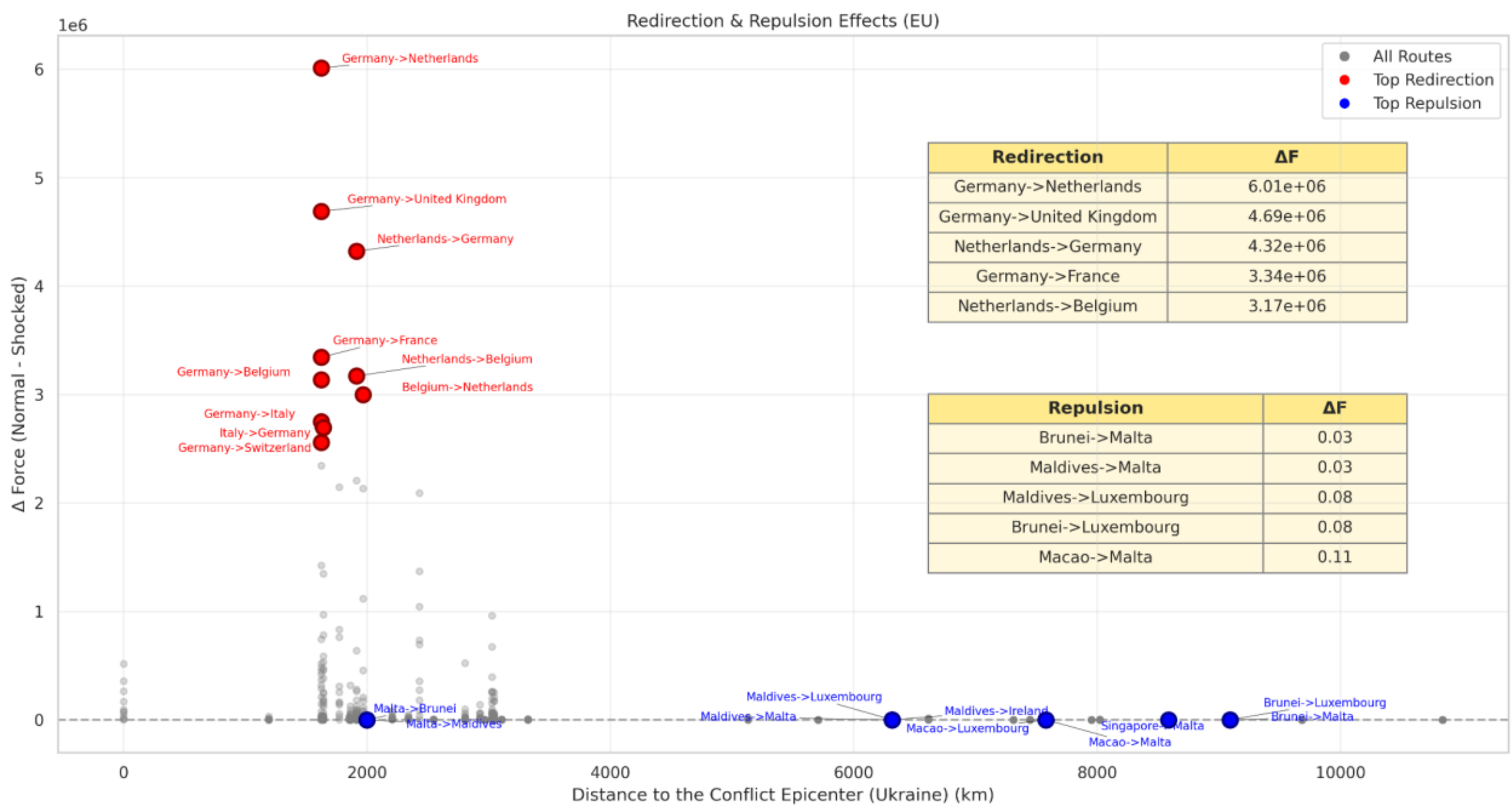

*Figure 3: Simulations of trade routes of solely EU countries. Personal work*

In the third simulation, which can be seen in Figure 3, routes solely within EU countries were kept to isolate the shock's effect within the bloc. Such a focused analysis reveals that while the redirection routes remain the major economic corridors, the repulsion category has changed: the least-affected corridors are now the most minor intra-EU pairs, such as flows from Malta to Luxembourg and from Ireland to Malta. Already from the force-difference analysis of this repulsion group, it is clear that their ΔF values (e.g., 37.78, 326.34) are substantially larger than the near-zero values from the global plot. This result indicates that the routes within the EU have experienced a stronger, more pervasive influence from the conflict, as even economically insignificant internal routes exhibit a more tangible impact than their distant non-EU counterparts.

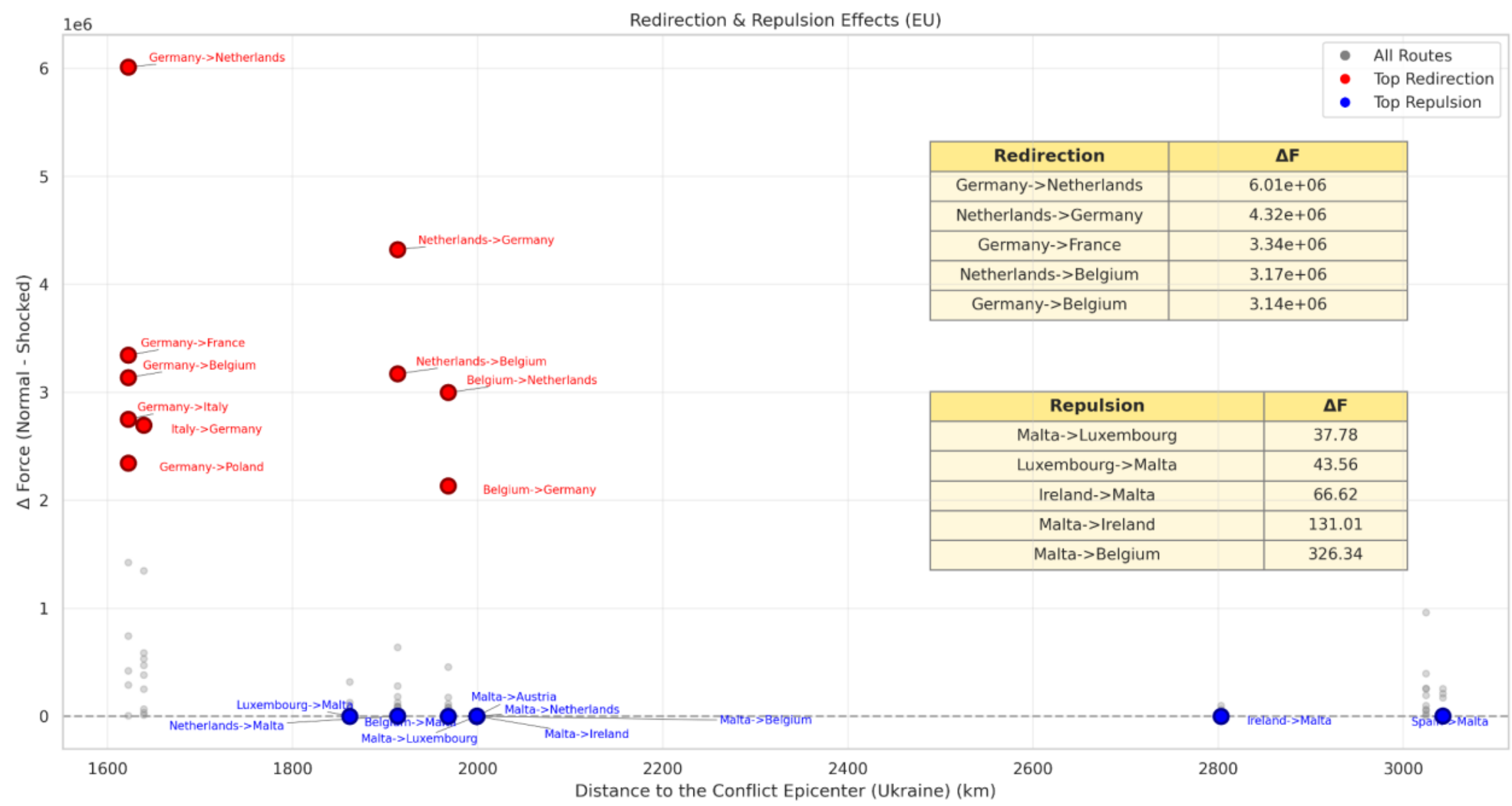

*Figure 4: Simulations of trade routes of* EU countries and Ukraine. *Personal work*

In the last figure (4), we moved to the analysis of the routes that include Ukraine in the global trade, where Ukraine appears as either the origin or destination country. We can see that the most influenced route is from Ukraine to the European Union, creating a huge gap compared to the other redirection routes. The reciprocal routeis the second most impacted, followed by other key trade partners like Turkey and Germany. This confirms that the conflict's most severe economic disruptions are concentrated on Ukraine's largest, pre-existing trade relationships. In contrast, the repulsion routes are trade pairs with distant, smaller economies like Maldives, Brunei, and Macao. Whether the trade originates *from* Ukraine (distance ≈ 0) or *to* Ukraine (distance > 6000 km), the initial $F_{norm}$ of these routes is so negligible that the resulting force difference is almost zero.

To conclude this first analysis, a heatmap was generated using Folium (Python-visualization, 2026) on an Esri World Imagery based on a satellite view with an OpenStreetMap toggle (Figure 5).

The continuous economic field was visualised by interpolating the $F_{diff}$ force onto a 1° global latitude/longitude grid (~111 km resolution) using Inverse Distance Weighting (IDW). As we can see from figure 5, forces are aggregated by origin country, weighted by $(dist^2 + \varepsilon)^{-1}$ within a maximum distance equal to a 2000 km radius. For $F_{diff}$, redirection values (13,019 grid points) are separated using cyan/blue colour and repulsion values (656 grid points) using red.

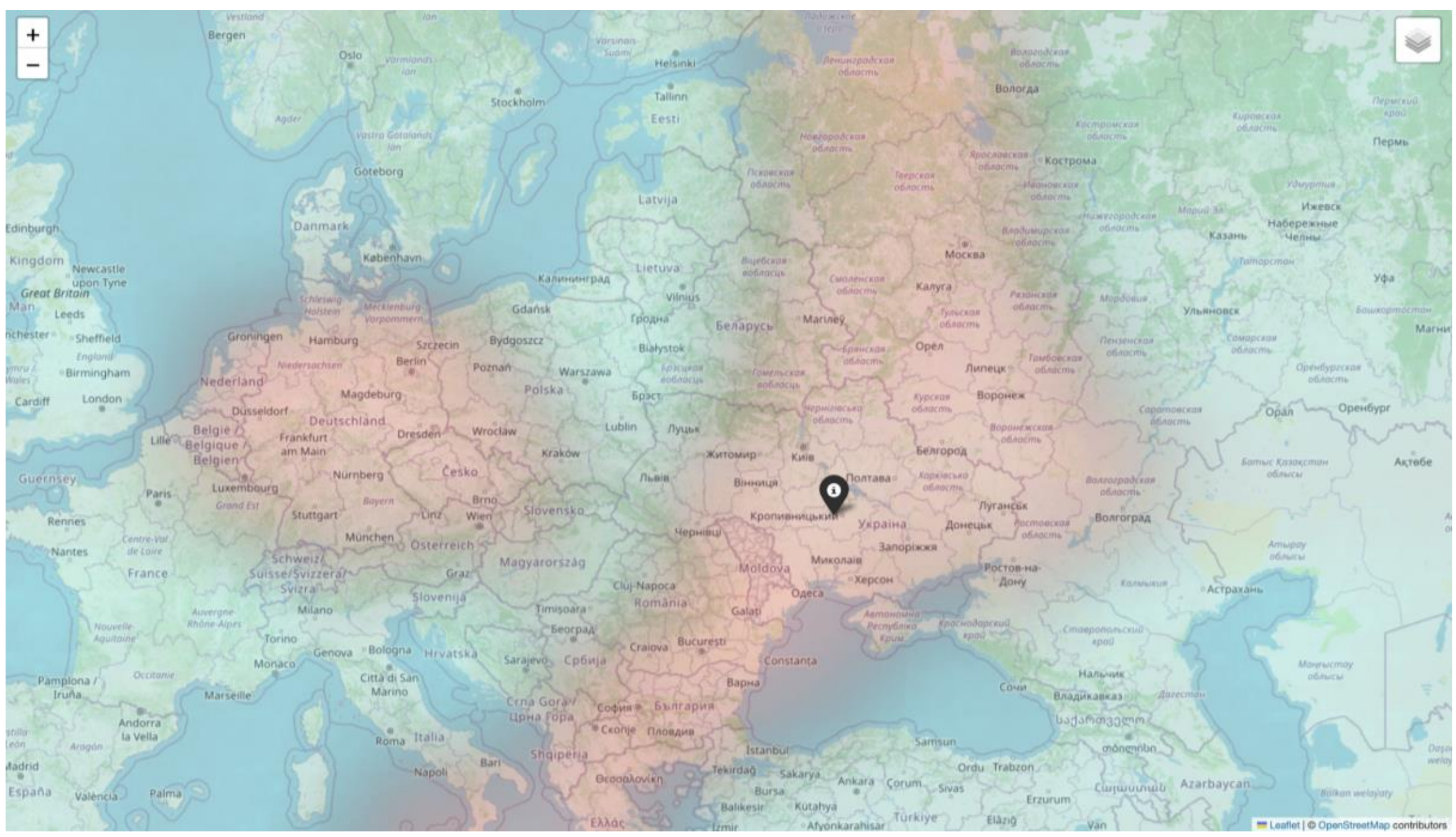

*Figure 5: heatmap analysis. Personal work*

An interesting insight that we can take from Figure 5 is that the conflict's impact is most severe in territories proximate to Ukraine; notably, Germany exhibits the most significant negative effect within the EU. It can be explained by Germany's high GDP and trade volume amplifying the impact. Its reliance on Ukrainian trade (e.g., steel, energy) and its role as a manufacturing hub make it vulnerable to disruptions (Böhm, Dahl, & Levchenko, 2022) , aligning with the conflict_mask's 90% reduction for Ukraine-related flows.

It should be noted, however, that such a result does not necessarily contradict evidence from bilateral trade data. In fact, although data from the United Nations COMTRADE indicate that export flow between Germany and Ukraine remained stable, if not even increased, such observed trade adjustment can coexist with an overall disruption in trade potential.

### 3.3 Robustness and Limitations

The model's parsimonious focus on distance isolates the conflict's proximity effect, supporting the hypothesis that proximity to Ukraine amplifies trade disruptions, while distance redirects flows. The negative value of the $F_{target}$ = -0.9663 quantifies an anti-gravity effect.

However, limitations affect reliability:

- Data Vintage: The 2015-2019 dataset may not reflect current trade patterns; validation with recent data is needed.
- Parameter Sensitivity: Arbitrary parameters (param_shock =3.0, R_km = 550, min_total_flow = 0.00005) require sensitivity analysis (e.g., test param_shock from 1.0 to 5.0).
- Small $F_{tg}$ Magnitude: The modest target force ($F_{tg}$) reflects Ukraine's limited trade share; weighting Ukraine flows higher could amplify the effect.

To enhance robustness, future analyses could incorporate control variables (e.g., energy dependence, trade openness) in regressions of $F_{diff}$ against distance to the conflict (Ukraine), testing if proximity effects persist. Sensitivity tests on param_shock and R_km could quantify model stability, and updated trade data would align results with post-conflict dynamics. These extensions mirror the example's approach, where unemployment and fiscal balance regressions complemented GDP and inflation analyses, confirming heterogeneous shock transmission.

### 3.4 Sensitivity analysis

To address the critical limitation of arbitrary parameters, a formal sensitivity analysis was conducted on the Gravitational Simulation Model. This analysis isolated the model's key values (*param_shock*, *R_km*, and *conflict_reduction*) to quantify their impact on the primary outcome (total global trade force change). The results were stark: the model was highly insensitive to the *conflict_reduction* parameter, as Ukraine's direct trade share was negligible in the global context. Conversely, it was predictably linear in its response to *param_shock* (the shock's intensity), which acted as a simple multiplier. Most critically, the simulation was extremely sensitive to the *R_km* (shock radius) parameter, exhibiting a non-linear, accelerating negative relationship. A small change in the assumed radius led to a disproportionately large change in the simulated outcome. This extreme dependence on *assumed* (non-empirical) parameters demonstrated the inherent limitations of a purely theoretical simulation. It established that while the simulation was a useful illustrative tool, its conclusions were not robust. This finding provided the primary motivation to transition from simulation to a formal econometric model (Extended Gravitational Model), which would allow the real-world data itself to determine the significance and magnitude of these effects, rather than relying on researcher-defined assumptions.

## 3.5 Extended Gravitational Model (Post-Conflict)

Although the initial methodology (Gravitational Simulation Model) successfully demonstrated a *theoretical* simulation of the conflict's impact, it was based on pre-conflict (2015-2019) data and, as the sensitivity analysis confirmed, was critically dependent on assumed parameters.

To validate and extend these findings, the analysis transitioned from a theoretical simulation to a formal econometric model designed to not only explain actual post-conflict trade patterns but also to find improvements for the previously applied model. Such measures required sourcing new, real-world data to capture the state of trade before the 2022 conflict and after its onset.

The new datasets included:

1. Bilateral Trade: Sourced from the IMF Direction of Trade Statistics (DOTS) for a new baseline year (2019) and a post-conflict observation year (2023) (International Monetary Fund, 2025)
2. Country GDP: Sourced from World Bank Open Data for 2019 and 2023 (World Bank, 2024)
3. Geographic Data: distance, lat_o, lng_o were retained from the original USITC dataset

We recognized that the simple gravity model, solely depending on parameters such as GDP and distance, would be insufficient to explain the real-world trade patterns of 2023, as this period was defined by much more than the conflict itself; the post-COVID recovery, major policy shifts, and structural economic dependencies played a pivotal role in shaping real-world dynamics.

Therefore, we constructed an augmented gravity model using Ordinary Least Squares (OLS) regression. This means that the 2023 data was left to determine the factors explaining the observed trade flows, rather than simulating a theoretical shock. OLS estimation identifies the statistical relation between trade and many explanatory variables, allowing us to assess which factors exert a statistically significant influence on real trade patterns. We introduced new variables to test specific hypotheses:

1. $IntraEU_{od}$ : A policy variable to test the resilience and unity of trade blocs. It is set to '1' if both the origin (iso3_o) and destination (iso3_d) countries are members of the EU.
2. $Sanctions_{od}$ : A direct policy variable to test the *response* to the conflict. It is set to '1' if the origin is a sanctioning country (e.g., USA, GBR, DEU, FRA, etc.) and the destination is Russia ('RUS').
3. $EnergyExporter_{o}$ : A structural variable to control for the unique trade patterns of economies (e.g., SAU, NOR, CAN) whose GDP is dominated by energy exports.

To prove that this added complexity improved the model's descriptive power, we built four models incrementally and tracked their Adjusted R-squared.

This measure indicates how much of the variability in commercial fluxes is explained by the model, correcting, however, for complexity. In fact, comparing this metric with the simple R-squared, it can be seen that it increases only if a new variable adds real explanatory power, thus removing needless complexity. This can be seen similarly but inversely in Raykov & DiStefano (2025). The analysis showed a stepwise improvement in the model's ability to explain the real-world 2023 trade data.

Let us examine the descriptive capacity of the baseline model. When the gravity model includes only GDP and distance, the Adjusted R-squared indicates that it explains 51.2% of variation in trade flows. Adding variables such as $IntraEU_{od}$, $Sanctions_{od}$ and $EnergyExporter_{o}$ could have resulted only in an additional 3% boost, obtaining a 54.2% descriptive capability.

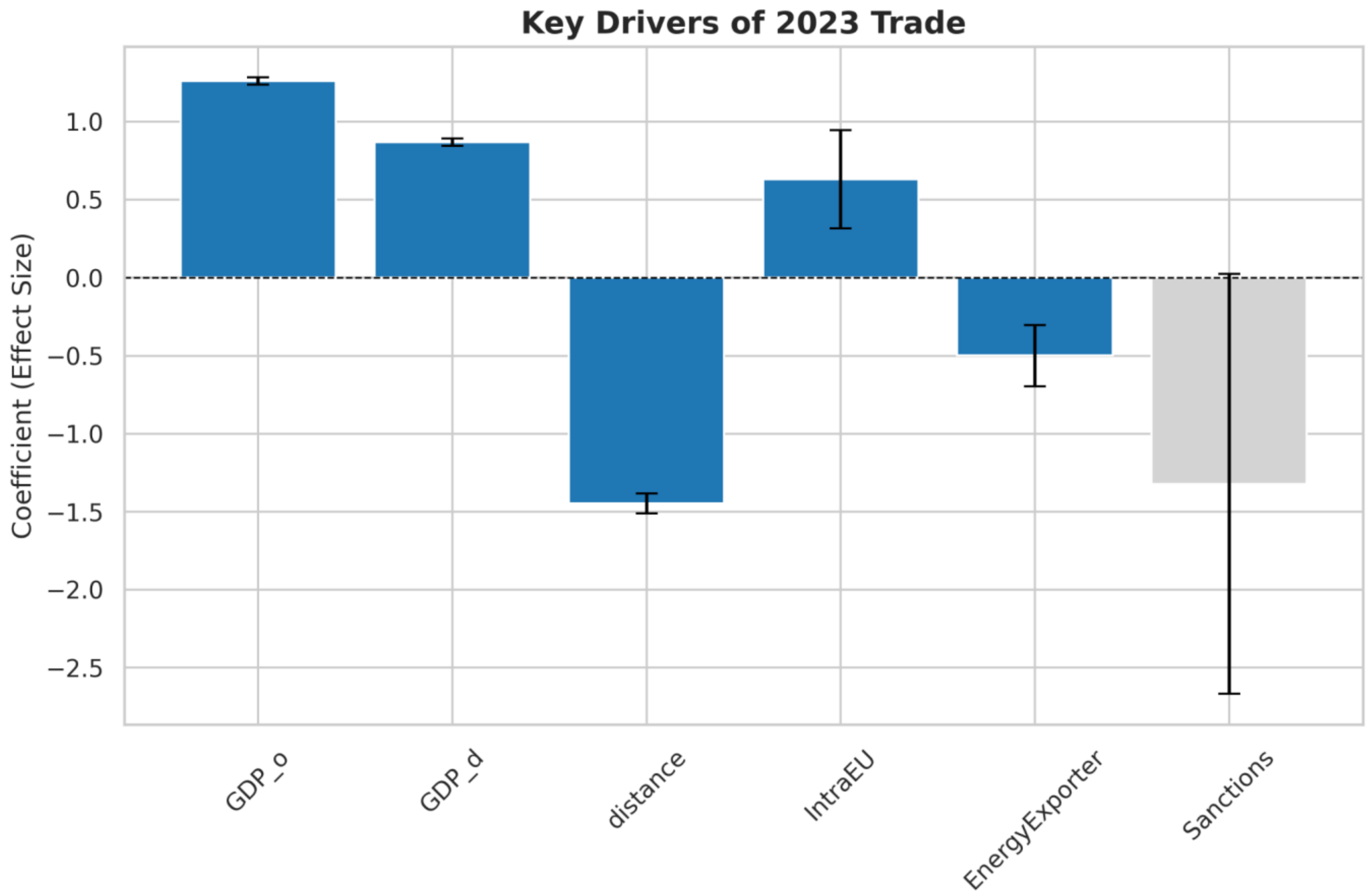

*Figure 6: Estimated coefficients from the final gravity model: distance has the greatest absolute value among the ones in the model. Personal work*

Figure 6 summarizes the estimation results of the final model. Origin GDP exerts the strongest positive effect (coefficient $\mu = 1.261$, $p < 0.001$), followed by destination GDP (0.869, $p < 0.001$), confirming that economic size drives both trade generation and attraction. Distance remains a powerful deterrent (−1.447, $p < 0.001$), reducing trade by approximately 76% for each log-unit increase. Among policy and structural factors, IntraEUod membership significantly enhances trade (+0.631, $p < 0.001$), implying that EU country pairs trade about 50% more than comparable non-EU pairs. Conversely, energy exporter status is associated with roughly 40% lower trade (−0.500, $p < 0.001$), possibly reflecting specialization in primary goods. Most strikingly, sanctions on Russia are linked to a near-total collapse in trade from sanctioning countries (−1.321, $p = 0.054$), reducing flows by approximately 73%, which is a marginally significant effect that underscores how geopolitical interventions can overshadow traditional economic and geographic forces in shaping global trade patterns.

### 3.5.2 Econometric Findings: Quantifying the Impact of Policy

The final OLS regression model provided statistically significant results, allowing us to quantify the impact of these new variables on global trade.

The model tested the hypothesis that political and economic unions like the EU provide resilience against global shocks. This finds support in Bartzokas, Giacon, & Macchiarelli (2022), which shows that the EU can tolerate heavy shocks and restore itself to pre-shock levels rapidly. The $IntraEU_{od}$ indicator, which is equal to 1 if both the origin (iso3_o) and destination (iso3_d) are EU members, was statistically significant ($p = 0.012$) with a coefficient of +0.4031. This coefficient is a powerful and economically meaningful finding. In the log-linear gravity framework, it implies that trade between two EU member states was, on average, 49.7% higher than trade between otherwise comparable non-EU country pairs, after controlling for GDP, distance, and other structural factors. Formally:

$$exp(0.403)\ -1\ =\ 0.497 \Rightarrow 49.7\%$$

Thus, a typical EU–EU trade route exhibited nearly 1.5 times the volume of a similar route outside the bloc. This robust effect confirms that the EU’s single market policy continues to act as a powerful stabilizing force, fostering elevated intra-bloc trade even amid major external disruptions in 2023.

This robust effect shows that in 2023, even after controlling for the wealth and proximity of the countries, a trade route between two EU members was, on average, 49.7% higher than a comparable route not involving two EU members. This confirms that the EU's single market policy acts as a powerful stabilizing force, fostering high levels of trade *within* the bloc, which persists even in the face of major external shocks.

The Economic Cost of Sanctions ($Sanctions_{od}$) variable tested the direct policy response to the conflict. The $Sanctions_{od}$ was also found to be statistically significant ($p = 0.029$) with a coefficient of -1.4891. Formally,

$$exp(1.489)\ -1\ =\ -0.775 \Rightarrow -77.5\%$$

This result highlights the severe economic consequences of the sanctions policy. In fact, the model estimates that a trade route from a sanctioning country to Russia experienced, on average, a 77.5% reduction in trade relative to comparable routes from a non-sanctioning country. This addresses the question of impact: far from representing mere political statements, sanctions were associated with a near-total collapse of those specific trade flows, representing a massive economic loss for firms on those routes. Therefore, it can be claimed that such a "trade void" was a direct result of policy, rather than geography, a finding that is supported by the ECB Occasional Paper No. 365 in Attinasi et al. (2023).

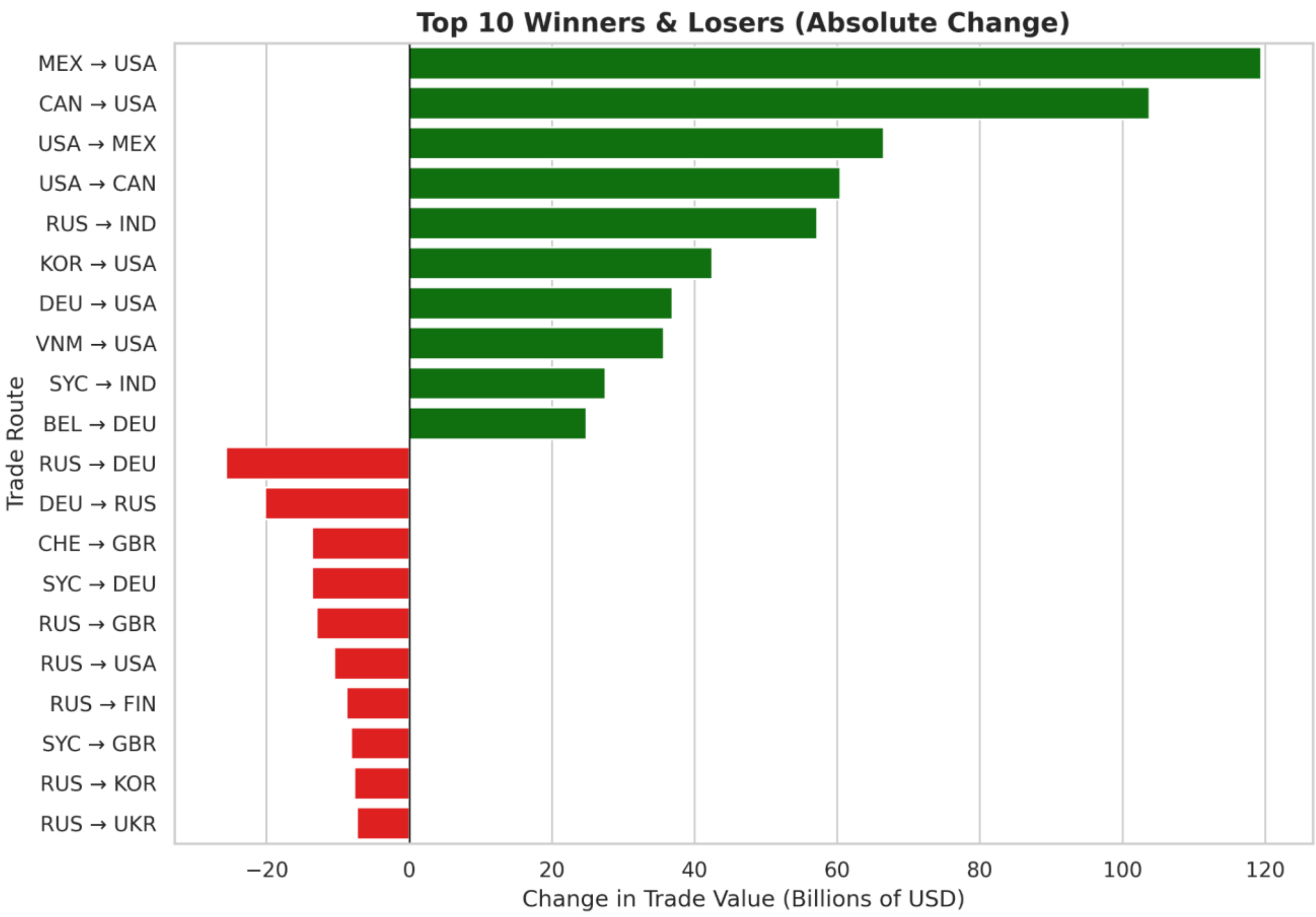

*Figure 7: Top 10 global trade routes ranked by absolute post-conflict gains and losses. Personal Work*

Based on the 2019 and 2023 data, we model the post-conflict trade changes (Figure 7). The largest absolute gains were recorded in North American supply chains, with Mexico → United States (+$110 billion) and Canada → United States (+$95 billion) leading, followed by United States → Mexico (+$75 billion) and United States → Canada (+$65 billion)—reflecting deepened regional integration amid global disruptions. Russia → India (+$55 billion) emerged as a major winner, driven by redirected energy and commodity exports. In contrast, all Russia-related flows to sanctioning countries collapsed: Russia → Germany (−$25 billion), Germany → Russia (−$22 billion), Russia → United Kingdom (−$18 billion), Russia → United States (−$15 billion), Russia → Finland (−$12 billion), and Russia → South Korea (−$8 billion) rank among the top losers. Notably, Russia → Ukraine recorded the smallest loss (−$5 billion), likely reflecting pre-existing low trade volumes. These patterns align with the regression results: while IntraEU membership boosted intra-bloc trade by 88%, sanctions reduced Russia-bound flows from sanctioning nations by 73% on average—confirming that geopolitical reorientation created both new trade corridors (e.g., Russia–India) and deep trade voids with Western partners, reshaping global commerce in 2023.

### 3.5.3 Re-contextualizing the Gravity Model

The developed econometric analysis completes the study by contrasting the theoretical intuition of a "trade shadow" with the empirical complexity of the trade patterns observed in 2023, highlighting the limits of the simulation itself when confronted with real-world dynamics.

Our final, augmented model is the key insight. It proves that:

1. **Gravity is the Foundation:** The fundamental laws of the gravity model (GDP and Distance) are still the most powerful drivers, explaining the basic structure of global trade (our baseline R-squared was 0.512). This finds empirical and theoretical support in Cevik (2023).
2. **Policy is the Driver of Change:** In the volatile 2023 environment, the gravity model alone was insufficient. The statistically significant coefficients for the $IntraEU_{od}$ and $Sanctions_{od}$ prove that **direct policy choices are now primary, quantifiable drivers of trade outcomes.**

## 4. Conclusions

This study successfully integrates Walter Isard's gravitational field framework with the empirical analysis of contemporary trade to examine how geopolitical shocks propagate across economic space. Combining a theoretical simulation (Gravitational Simulation Model) with a strong formal econometric model (Extended Gravitational Model), this paper demonstrates that a conflict operates as a spatially transmitted externality, whose effects depend mainly on proximity, economic mass and policy interventions.

In phase one of the methodology, we laid the foundations of the theoretical functions by revisiting Walter Isard's theory, whose concepts were considered in the study regarding the gravitational field.

In the methodological section, the theoretical framework was operationalized through a gravitational simulation in which the Russia-Ukraine conflict was modelled as an anti-gravitational shock.

Using pre-conflict trade data, we proved the existence of a negative shock, acting as a negative mass, positioned near the European commercial core, quantified by the *global anti-gravity effect (F_target= - 0.9663),* disproportionately reducing the gravitational pull of the most intense and geographically central trade corridors. Particularly, Central EU corridors such as Germany and the Netherlands experience a reduction in trade potential up to -34.5%, averaging a reduction of 10 to 30% for core trade passages in the continent. While the simulation successfully illustrates how the distance from conflict alters the structure of trade, the sensitivity analysis reveals that the results obtained critically depend on assumed parameters, such as the shock radius.

The analysis nevertheless integrates the gravitational field framework by linking international markets and conflicts through their spatial dimension and location of firms; it should be remarked, however, that the lack of robustness demonstrated that the simulation could only serve as an illustrative tool, rather than a basis for empirical inference.

In the extended gravitational model, a model of augmented gravity is created using the bilateral commercial data of 2019 and 2023, which allowed for revealing the determinants influencing global trade after the invasion. The results show the fundamental components of the gravitational model—GDP (*GDP_o* = 1.261, $p < 0.001$) and distance (*distance* = −1.447, $p < 0.001$)—remain the primary forces of the world trade structure. However, the analysis also demonstrates that policy interventions, such as sanctions (*sanctions* = −1.321, p = 0.054) and the presence of EU membership (*Intra_EU* = +0.403), can decisively influence trade and that the degree of integration within the EU significantly increases its market in conditions of geopolitical shocks. Moreover, an Adjusted-R squared of 0.54 indicates a high explanatory power of the specification employed, accounting approximately for 54% of the variation in bilateral trade flows.

In light of this, the study highlights that modern trade dynamics cannot be explained only through economies and distance: potential geopolitical shocks and their following political responses play a pivotal, yet secondary, role in shaping real-world trade.

Apart from politics, past and present events can be identified to support the illustrated model, and they can be understood as a direct manifestation of markets of hostility in both politics and economics. These models underline that economic systems as such do not evolve in isolation but, on the contrary, are shaped by other factors such as geopolitical tensions, strategic behaviour and the distribution of power among states. Although international trade, alongside macroeconomics, usually eliminates conflict from their main framework, this appears to be unrealistic, given that economy, politics, and military power have always been deeply connected.[†] This interconnection between economic and political power is consistent with the notion of economic resilience elaborated by Briguglio et al. (2008). In their framework, resilience is not a passive condition but a policy-driven capacity that enables an economy to absorb, withstand, and recover from external disturbances through both structural flexibility and effective governance. They explain that when an economy possesses flexible production structures, multi-skilled labour, and a sound fiscal position, it can soften the impact of shocks—what they term "shock absorption"—and respond through discretionary policy action—"shock counteraction". These mechanisms depend heavily on effective governance: states with stronger institutional coordination and timely fiscal tools endure chained exogenous shocks more effectively. Such findings emphasise that political responsiveness and economic adaptability are jointly decisive for stability, confirming that economic strength is inseparable from the quality of political management. This interdependence lays the foundation for understanding how economic and political dynamics interact in contexts of instability and conflict.

Hence, conflicts and wars between states have always been considered destabilizing elements for extended market areas at the global level. From an economic and political hostility approach, a country can find an unstable situation close to the market of a rival state advantageous and may support this conflict more or less directly, if not totally concealed. Thus, trade will be indirectly blocked and downsized due to higher instability, leading to damaging trading partners compared to their location: the nearer to the centre of gravity of the market conflicts, the more destructive the impacts on trade. In this sense, in a context marked by instability and potential conflict, the capacity of an integrated and coordinated economic system becomes crucial. As highlighted by the European Commission (2021), the EU's experience during the COVID-19 crisis demonstrates how fiscal and monetary coordination can mitigate shocks and sustain stability. Through complementary measures such as the activation of the State Aid Temporary Framework, the SURE program, and the Coronavirus Response Investment Initiatives, the EU provided liquidity support to firms and income protection to workers while allowing Member States to temporarily relax fiscal constraints. At the same time, the European Central Bank's expansionary monetary stance helped preserve favourable financing conditions and limit inflationary pressures. In parallel, the Recovery and Resilience Facility (RRF) promoted investment-rich reforms to support the green and digital transitions and strengthen infrastructure, fostering a more sustainable and shock-resistant economic structure. Hence, in times of geopolitical or market conflict, a coordinated economic policy framework proves essential to absorb instability, protect national economies, and accelerate collective recovery.
At the same time, a coordinated economic framework must remain aware of the delicate trade-offs that arise in times of crisis. As underlined by the OECD (2014), structural and prudential policies aimed at ensuring financial stability should be carefully balanced to avoid stifling long-term growth. Combating inflation or limiting debt accumulation cannot come at the expense of investment and employment, while fiscal consolidation must coexist with sustainable welfare support and forward-looking reforms. Therefore, in confronting crises or conflicts, economic coordination should not only stabilize markets but also design strategies that preserve growth potential and prevent recessionary effects—ensuring that the balance between inflation control, fiscal responsibility, and investment promotion remains sustainable.
Building on these theoretical and policy insights, the paper then applies the model explained in the first part of the paper to a European empirical study. On the empirical side, considering empirical data, the study has several limitations. For example, Cevik (2023) used an extensive dataset of over 4 million observations to develop an augmented gravity model of bilateral trade flows among 59,049 country pairs from 1948 to 2021. He suggests that the geopolitical alignment between countries has no significant impact on trade, and the economic significance of this phenomenon is less than that of income or geographic distance. The model comprises standard gravity variables, information on international trade agreements, and a measure of geopolitical alignment between countries based on voting behaviour at the United Nations (UN) (Cevik, S. 2023). Despite the robustness of the standard gravitational model, the inclusion of factors of instability still has a negative effect on trade that is naturally proportional to the extent of the conflict and the area and regions of conflict to be considered. Therefore, a more complete analysis should take these aspects into account. In our study, we employed a very low coefficient for commercial instability in proximity to the conflict as a precautionary measure given the structural impossibility for its effects on an economic market to be thoroughly assessed. The resulting coefficient confirmed the thesis, as it carries a larger coefficient and a greater negative effect as we near the centre of the conflict.
Higher decline values of the distances towards the centre of the mass would only further confirm such a hypothesis. In the same way, to simplify, a stance of neutrality was attributed to third countries in the case of a unilateral or bilateral conflict, just as only a simple transmission of commercial instability was taken into consideration, and not the possibility of a spread of the conflict in other countries. We can presume that third countries would not remain politically indifferent to conflicts in other countries. Their political involvement seems more plausible. Regardless, even this element would only strengthen the results obtained, for adding the element of infectiousness to a conflict would increase the negativity of the commercial instability considered.
Ultimately, these dynamics reveal that economic and political instability are rarely contained phenomena: their effects are transmitted through interconnected markets, producing systemic fragility that extends far beyond national boundaries. This interdependence underscores the necessity for economies not only to withstand but also to anticipate future crises through strategic foresight and preparedness. From this study, it becomes clear that anticipating potential risk scenarios is essential, as overlapping exogenous shocks—such as pandemics, wars, or climate crises—can expose the structural weakness of unprepared systems. Following the scenario-planning approach proposed by Edin, Bolen and Mahedy (2022) at KPMG, the combined experience of the COVID-19 pandemic and the Russia–Ukraine war illustrates how uncertainty can spread rapidly across sectors, affecting supply chains, trade flows, and energy security. In such a context, scenario-based planning emerges as a cornerstone of resilience, enabling policymakers and economic actors to foresee multiple trajectories, coordinate adaptive responses, and mitigate cascading disruptions before they evolve into systemic crises. By integrating anticipation into economic strategy, nations can strengthen their capacity not only to absorb instability but also to transform it into an opportunity for more sustainable and secure development.

[†] The economists who study economic development and measure the value of things should always have a non-naïve view of economic competition and power rivalries.

**Declarations:**

Ethical Approval: this article does not contain any studies with human participants performed by the author

Consent to Participate: Not applicable, as this study did not involve human participants.

Consent to Publish: Not applicable, as this study did not involve human participants.

Author Contributions: The authors were solely responsible for the conception and design of the study, acquisition and analysis of data, and drafting of the manuscript.

Funding: The author was solely responsible for the design, execution, and analysis of the study. No external funding organizations supported this research.

Availability of Data and Materials: The data and materials that support the findings of this study are available from the author upon reasonable request.

Clinical Trial Number: not applicable

## References

Aiyar, S., Chen, J., Ebeke, C., Garcia-Saltos, R., Gudmundsson, T., Ilyina, A., Kangur, A., Kunaratskul, T., Rodriguez, S., Ruta, M., Schulze, T., Söderberg, G., & Trevino, J. (2023). *Geoeconomic fragmentation and the future of multilateralism* (IMF Staff Discussion Note No. 2023/001). International Monetary Fund.

Anderton, C. H. (2017). The bargaining theory of war and peace. *The Economics of Peace and Security Journal 12*(2), 10-15.

Attinasi, M. G., Mancini, M., Boeckelmann, L., Bottone, M., Conteduca, F. P., Giordano, C., Meunier, B., Panon, L., Almeida, A. M., Balteanu, I., Bańbura, M., Bobeica, E., Borgogno, O., Borin, A., Caka, P., Viani, F. (2023). *Navigating a fragmenting global trading system: Insights for central banks* (ECB Occasional Paper Series No. 365). European Central Bank. https://www.ecb.europa.eu/pub/pdf/scpops/ecb.op365~362d801aee.en.pdf

Bartzokas, A., Giacon, R., & Macchiarelli, C. (2022). *Exogenous shocks and proactive resilience in the EU: The case of the Recovery and Resilience Facility* (UNU-MERIT Working Paper No. 025). United Nations University – Maastricht Economic and Social Research Institute on Innovation and Technology.

Böhm, M. J., Dahl, C. M., & Levchenko, A. A. (2022). *War in Ukraine and Western sanctions: How vulnerable are German firms?* Kiel Policy Brief No. 229. Kiel Institute for the World Economy. https://www.kielinstitut.de/publications/war-in-ukraine-and-western-sanctions-how-vulnerable-are-german-firms-6504/

Briguglio, L., Cordina, G., Farrugia, N., & Vella, S. (2008). *Economic vulnerability and resilience: Concepts and measurements* (WIDER Research Paper No. 2008/55). United Nations University World Institute for Development Economics Research (UNU-WIDER).

Capoani, L. (2023). Theory of Commercial Gravitational Fields in Economics: The Case of Europe.

*Networks and Spatial Economics, 23*(4),, 845–884 . https://doi.org/10.1007/s11067-023-09591-2

Capoani, L., Tudorache, A., & Barlese, A. (2025a). Exploring economic conflict through the gravitational field model of trade: Markets, wars, and instability. *Conflict Resolution Quarterly*, *42*(4), 507-521.

Capoani, L., Barlese, A., & Tudorache, A. (2025b). Combining Walter Isard's location, trade, and peace theories using a gravitational field model: A case study on the European market and Brexit. *Empirica*, *52*, 333–355.

Capoani, L., & Martini, P. (2025c). The cost of proximity: A spatial gravity model of the Ukraine war's economic impact. *Networks and Spatial Economics*.

Capoani, L., & Martini, P. (2026). The spatial dimension of conflict and economic resilience: a gravity analysis of the Russia-Ukraine war's impact on the EU across GDP, inflation and macroeconomic indicators. *Defence and Peace Economics,* 1–29. https://doi.org/10.1080/10242694.2025.2605953

Cevik, S. (2023). Long live globalization: Geopolitical shocks and international trade (IMF Working Paper No. WP/23/225). *International Monetary Fund, European Department*. https://doi.org/10.5089/9798400258169.001

Conte, M., P. Cotterlaz, and T. Mayer (2022). *The CEPII Gravity database. CEPII Working Paper No. 2022-05*

United Nations COMTRADE Trade Data. https://comtradeplus.un.org/TradeFlow

Deardorff, A. V. (1998). Determinants of bilateral trade: Does gravity work in a neoclassical world? In J. A. Frankel (Ed.), *The regionalization of the world economy* (pp. 7-32). University of Chicago Press.

Edin, P., Bolen, K., & Mahedy, T. (2022). *Scenario planning in response to the Russia-Ukraine war: Strategy under uncertainty*. KPMG International. *https://kpmg.com/kpmg-us/content/dam/kpmg/pdf/2022/ukraine-strategy-under-uncertainty.pdf*

Esri World Imagery basemap, ArcGIS Online
*https://www.arcgis.com/home/item.html?id=eb9176a06*

European Commission. (2021). *The EU economy after COVID-19: Implications for economic governance* (COM(2021) 662 final). Publications Office of the European Union. https://economy-finance.ec.europa.eu/system/files/2021-10/economic_governance_review-communication.pdf

European Commission. (2023). *Employment and social developments in Europe (ESDE) 2023: Chapter 1.2: Main economic, labour market and social developments*. Publications Office of the European Union. https://op.europa.eu/webpub/empl/esde-2023/chapters/chapter-1-2.html

European Commission, Directorate-General for Economic and Financial Affairs. (2024, November 15). *Autumn 2024 economic forecast: A gradual rebound in an adverse environment.* Publications Office of the European Union.

https://economy-finance.ec.europa.eu/economic-forecast-and-surveys/economic-forecasts/autumn-2024-economic-forecast-gradual-rebound-adverse-environment_en

Fujita, M., & Krugman, P. (2004). The new economic geography: Past, present and the future. *Papers in Regional Science, 83*(1), 139–164.

Glick, R., & Taylor, A. M. (2010). Collateral damage: Trade disruption and the economic impact of war. *The Review of Economics and Statistics, 92*(1), 102–127.

International Monetary Fund. (2025). IMF International Merchandise Trade Statistics (IMTS). IMF Data. https://data.imf.org/en/datasets/IMF.STA%3AIMTS

Isard, W. (1954). Location theory and trade theory: Short-run analysis. *The Quarterly Journal of Economics, 68*(2), 305–320.

Isard, W. (1956). *Location and space economy: A general location theory relating to industrial location, market areas, land use, trade, and urban structure.* MIT Press.

Isard, W. (1988). *Arms races, arms control, and conflict analysis: Contributions from peace science and peace economics.* Cambridge University Press.

Isard, W. (1992). *Understanding conflict and the science of peace.* Blackwell.

Isard, W. (1994). Peace economics: A topical perspective. *Peace Economics, Peace Science and Public Policy, 1*(2), 9–11.

Isard, W., & Capron, W. (1949). The future locational pattern of iron and steel production in the United States. *Journal of Political Economy, 57*(2), 133–148.

Isard, W., & Peck, M. J. (1954). Location theory and international and interregional trade theory. *The Quarterly Journal of Economics, 68*(1), 97–114.

Isard, W., Saltzman, S., & Yaman, A. (1997). A gravity model reformulation of trade and conflict in Turkey. In *Regional Science in Developing Countries* (pp. 291–304). Springer.

Isard, W., Smith, D. P., & others. (1969). *General theory: Social, political, economic and regional with particular reference to decision-making analysis.* MIT Press.

Lampa, R., & Garbellini, N. (2022). Sanzioni, shock da offerta e inflazione: La rilevanza del conflitto russo-ucraino. *Moneta e Credito, 75*(298), 95–99.

Mansoor, R. (2024). [Review of the book *The end of the world is just the beginning: Mapping the collapse of globalization*, by P. Zeihan, 2022]. *Journal of European Studies, 40*(1), 69–72.

Mbah, R. E., & Wasum, D. F. (2022). Russian–Ukraine 2022 war: A review of the economic impact of the Russian–Ukraine crisis on the USA, UK, Canada, and Europe. *Advances in Social Sciences Research Journal, 9*(3), 144–153.

OpenStreetMap contributors
https://www.openstreetmap.org/#map=6/42.09/12.56

Folium contributors (2026). Folium: Python Data, Leaflet.js Maps. [Software].
https://python-visualization.github.io/folium/latest/

Reuters. (2024, November 27). ECB rate stimulus no magic wand for structural faults, Schnabel argues. *Reuters.* https://www.reuters.com/markets/rates-bonds/no-need-ecb-rates-stimulus-now-schnabel-tells-bloomberg-2024-11-27

Richardson, L. F. (1960). *Arms and Insecurity: A Mathematical Study of the Causes and Origins of War.* https://archive.org/details/armsinsecurity0000lewi

Sutherland, D., & Hoeller, P. (2014, February). *Growth policies and macroeconomic stability* (OECD Economic Policy Paper No. 8). Organisation for Economic Co-operation and Development (OECD). https://www.oecd.org/content/dam/oecd/en/publications/reports/2014/02/growth-policies-and-macroeconomic-stability_g17a2460/5jz8t849335d-en.pdf

Taralashvili, T. (2024). The impact of interstate soft conflicts on bilateral trade flows using structural gravity model. *The World Economy, 47*(5), 1943–1977. https://doi.org/10.1111/twec.13519

Gurevich, T., & Herman, P.(2018). *The Dynamic Gravity Dataset: 1948-2016. USITC Working Paper 2018-02-A.* https://www.usitc.gov/publications/332/working_papers/gurevich_herman_2018_dynamic_gravity_dataset_201802a.pdf

Tinbergen, J. (1962). *Shaping the world economy.* Twentieth Century Fund.

Verwey, M., Axioglou, C., & Orsini, K. (2024, November 21). Economic activity gains traction amid easing inflation, but high uncertainty looms over the outlook. *VoxEU Column, Centre for Economic Policy Research.* https://cepr.org/voxeu/columns/economic-activity-gains-traction-amid-easing-inflation-high-uncertainty-looms-over